\documentclass{ws-p9-75x6-50}

\renewcommand{\S}{\hat S}
\newcommand{\BEQ}{\begin{equation}}
\newcommand{\EEQ}{\end{equation}}
\newcommand{\BEA}{\begin{eqnarray}}
\newcommand{\EEA}{\end{eqnarray}}

\renewcommand{\d}{{\rm d }}

\renewcommand{\S}{S_{\rm ep}}
\newcommand{\p}{\partial}

\renewcommand{\S}{S_{\rm ep}}
\newcommand{\Tbg}{T_{ CMB}}
\renewcommand{\TH}{T_H}

\newcommand{\I}{{\cal I}}
\newcommand{\bi}{\bibitem}

\def\dbarrm {{\mathchar'26\mkern-11mu{\rm d}}}

\newcommand{\fix}{{\Bigl.\Bigr|}}

\begin{document}

\title{Thermodynamics Far from Equilibrium: from Glasses to Black Holes}

\author{Th.M. Nieuwenhuizen}
\address{Department of Physics and Astronomy,\\
Valckenierstraat 65, 1018 XE Amsterdam, The Netherlands\\
E-mail:nieuwenh@science.uva.nl}

\maketitle

\abstracts{
A framework for the non-equilibrium thermodynamics of glasses
is discussed. It also explains the non-equilibrium thermodynamics
of a black hole isolated from matter. The first and second laws 
of black dynamics and black hole thermodynamics are shown to coincide,
while the third laws deal with different physical issues.}

\section{Introduction}

Thermodynamics is the old science that describes the flow of energy in
systems with many atoms. Works started by  Carnot, Kelvin and
Clausius showed that these laws are very general. 
This universality led to the formulation of the first and second 
law of thermodynamics, that apply to a vast amount of systems, 
such as gases and crystals.

During half a century
there was still a problem with the application to glasses.
In this field there were classical paradoxes related to the 
so-called Ehrenfest relations and Prigogine-Defay ratio.
The solution of this problem is discussed below.
Due to its inherent non-equilibrium  nature a glass is far from
equilibrium. To describe it in a thermodynamic treatment 
one has to take into account
at least one additional system parameter, 
the self-generated effective temperature, and its conjugate variable, 
the configurational entropy ~\cite{Nthermo}\cite{Nhammer}.

After having realized how thermodynamics should be formulated for
glasses, we have investigated the situation for black
holes~\cite{Nblackhole}. For this problem
various aspects of the dynamics are known, and it was
generally expected that the laws for black hole dynamics 
would coincide with the laws of black hole thermodynamics.
We nevertheless felt that the proper connection
between black hole dynamics and standard thermodynamics had not been
made, and this will be clarified below.

Two years have passed since our letter on the black hole
thermodynamics appeared
~\cite{Nblackhole}. Till now the reaction of workers in 
gravitation was reserved, though researchers in other 
areas were attracted by its unifying concept.  
We feel that the blame should be put on the lack
of basic thermodynamic training in physics teaching programs.
Indeed, for many scholars thermodynamics is not much more than
Gibbsian equilibrium thermodynamics. If asked ``what is the second
law?'' the answer is often $\d U=T\d S-p\d V$, which, however, is the
combination of the first law ($\d U=\dbarrm Q+\dbarrm W$ with 
$\dbarrm W=-p\d V$ for a fluid) and the second law in case of
equilibrium, where the heat added to the system satisfies 
$\dbarrm Q=T\d S$. When mentioning the issue of
thermodynamics of a black hole isolated from matter to people
working in gravitation, a standard reaction is: 
``But I can put it in an insulating box, and then I can apply
equilibrium thermodynamics''. To us it is worrisome that such 
an unrealistic setup is considered to be a satisfactory explanation of
the issue. Of course, what is at stake is the non-equilibrium
thermodynamics of the black hole in the presence of cosmic background
radiation, and it is found that it is not much
different from the one for a black body. Even the question 
``Would you also claim that for a Newtonian star the background 
radiation should be taken into account?'' has a simple answer:
``Yes, in principle it should, but since the surface temperature 
  is then much larger, one could say that the outside 
has zero temperature''. And this statement is just equivalent
to saying that, by assumption, the Newtonian star is isolated, 
which makes it different from a star immersed in a gas cloud. 
Conceptually it is also important that requiring the outside 
to have (practically) zero temperature also excludes the situation 
where only some interior part of a star is discussed. 
So, for obvious reasons, the picture is self-consistent.

In short, we wish to stress that thermodynamics is not some
nasty issue that can be best avoided, rather it embodies the heart of
the physics under consideration, also, within appropriate 
time-windows, for a wide range of problems in the field of 
gravitation, such as black holes, Newtonian stars, globular star clusters, etc. 

To explain the issue we shall first consider the non-equilibrium 
thermodynamics of glasses, and then apply those insights to 
show that known results for isolated black holes perfectly fit within
this framework.

\section{Glasses}

Thermodynamics for systems far from equilibrium
has long been a field of confusion. A typical application is window glass.
Such a system is far from equilibrium: a cubic micron of glass is neither
a crystal nor an ordinary under-cooled liquid.
It is an under-cooled liquid that, in the glass formation process,
has fallen out of its meta-stable liquid equilibrium. There is 
thus a separation of timescales between the fast processes, often called
$\beta$-processes, and the basically quenched $\alpha$- 
or configurational processes.

Until our recent works on this field, the general consensus reached
after more than half a century of research was:
{\it Thermodynamics does not work for glasses,
because there is no equilibrium}. 
This conclusion was mainly based on
the failure to understand the Ehrenfest relations
and the  related Prigogine-Defay ratio. It should be kept in mind
that, so far, the approaches leaned very much on equilibrium ideas.
~\cite{DaviesJones}{}~\cite{GibbsDiMarzio}~\cite{DiMarzio1981}.
We shall stress that such approaches are not applicable, due to the
inherent non-equilibrium character of the glassy state.
Even the quoted conclusion itself is actually confusing, since
thermodynamics should also hold outside equilibrium.

Thermodynamics is the most robust field of physics.
Its failure to describe the glassy state is  quite unsatisfactory, 
since up to 25  decades in time can be involved.
Naively we expect that each decade has its own dynamics, 
basically independent of the other ones. 
We have found support for this point
in models that can be solved exactly.
Thermodynamics then means a description of system properties
under smooth enough non-equilibrium conditions, where the dynamics
is slow enough to build a quasi-equilibrium state.

\subsection{Two-temperature thermodynamics}

A state that slowly relaxes to equilibrium is characterized by the
time elapsed so far, sometimes called ``age'' or ``waiting time''.
For glassy systems this is of special relevance.
We shall restrict to systems with one diverging time scale.
They are systems with first-order-type transitions,
with (smoothed out) discontinuous order parameter, 
though usually there is no latent heat.

The picture to be investigated in this work starts
by describing a non-equilibrium state characterized by three parameters,
namely $T,p$ and the age $t$. 
By exactly solving the dynamics of certain model systems
~\cite{Nthermo}\cite{NEhren}\cite{Ndirpol}\cite{Nlongthermo}, 
it was found that in the thermodynamics the non-equilibrium 
nature shows up only through an
{\it effective temperature} $T_e(t)$;
also numerical data on the slow evolution of a binary Lennard-Jones glass
were recently interpreted in terms of an effective temperature
\cite{Kobby}.
For a set of smoothly related cooling experiments $T_i(t)$
at  pressures $p_i$, one may express the effective
temperature as a continuous function: $T_{e,i}(t)$ $\to$ $T_e(T,p)$.
This sets a surface in $(T,T_e,p)$ space, that becomes multi-valued
if one first cools, and then heats. For covering the whole space one
needs to do many experiments, e.g., at different pressures
and different cooling rates.
The results should agree with findings from heating experiments
and aging experiments.
Thermodynamics amounts to giving differential relations between
observables at nearby points in this space.
In principle also an effective pressure could be needed.
If there are many long time scales, also several effective
temperatures would be needed. We shall restrict ourselves to the
simplest case of one effective temperature.

Of special importance is the thermodynamics of a thermal body at
temperature $T_2$ in a heat bath at temperature $T_1$. This could
apply to mundane situations such as a cup of coffee, or an ice-cream,
 in a room. There are also two entropies, $S_1$ and $S_2$. Notice that
there are also two time-scales: the time-scale for heat to leave the
cup is much larger than the time-scale for equilibrating that
heat in the room. It is this separation of time-scales 
that makes spontaneous difference in temperatures possible.
The change in heat of such a system obeys  
$\dbarrm Q=\dbarrm Q_1+\dbarrm Q_2\le T_1\d S_1+T_2\d S_2$.

A similar two-temperature approach proves to be 
relevant for glassy systems.
The known exact results on the thermodynamics of systems without 
currents can be summarized by the very same change in heat
~\cite{NEhren} \cite{Nthermo}
\BEQ \label{dQ=}
\dbarrm Q=T\d\S+T_e\d\I
\EEQ
where $\S$ is the entropy of the fast or {\bf e}quilibrium
{\bf p}rocesses ($\beta$-processes)
and $\I$ the configurational entropy of the slow or configurational
processes ($\alpha$-processes). This object is also known as
information entropy or complexity. 
Notice that it deviates from the standard definition~\cite{GibbsDiMarzio} 
that the configurational entropy $S_c$ is the entropy
of the glass minus the one of the vibrational modes of the crystal.
The latter definition also incorporates fast non-vibrational processes.

It is both surprising and
satisfactory that a glass can be described by the same general law.
If, in certain systems, also an effective pressure or field would be 
needed, then $\dbarrm Q$
is expected to keep the same form, but $\dbarrm W$ would change from its
standard value $-p\d V$ for liquids.
Such an extension could be needed to describe a larger class of systems.

For a glass forming liquid the first law $\d U=\dbarrm Q+\dbarrm W$
becomes
\BEA \label{thermoglassp}
\label{dUp=}\d U=T\d \S+T_e \d \I-p\d V
\EEA
It is appropriate to define the generalized free enthalpy
\BEA \label{Gp=} G&=&U-T\S-T_e \I+pV\EEA
This is not the standard form, since $T_e\neq T$. It satisfies
\BEA
\label{dGp=}\d G&=&-\S\d T-\I\d T_e+V\d p
\EEA

The total entropy is
\BEQ \label{Stot=}S=\S+\I \EEQ
The second law requires $\dbarrm Q\le T\d S$, leading to
\BEQ (T_e-T)\d\I\le 0, \EEQ
which merely says that heat goes from high to low temperatures.

Since $T_e=T_e(T,p)$, and both entropies are functions of $T$, $T_e$
and $p$, the expression (\ref{dQ=}) yields a specific heat
\BEA
C_p&=&\frac{\p Q}{\p T}\fix_p=
T(\frac{\p \S}{\p T}\fix_{T_e,p}+
\frac{\p \S}{\p T_e}\fix_{T,p}\frac{\p T_e}{\p T}\fix_p)
+T_e(\frac{\p \I}{\p T}\fix_{T_e,p}+
\frac{\p \I}{\p T_e}\fix_{T,p}\frac{\p T_e}{\p T}\fix_p)
\EEA
In the glass transition region all factors, except $\p_T T_e$,
are basically constant. This leads to
\BEA \label{CpTool}
C_p&=&C_1+C_2\frac{\p T_e}{\p T}\fix_p
\EEA
Precisely this form has been assumed half a century ago
by Tool~\cite{Tool}  as starting point for
the study of caloric behavior in the glass formation region,
and has often been used for the explanation of experiments
{}~\cite{DaviesJones}\cite{Jaeckle86}.
It is  a direct consequence of eq. (\ref{dQ=}).

\subsection{Modified Maxwell relation, 
Ehrenfest relation and Prigogine-Defay ratio}

For a smooth sequence of cooling procedures of a
glassy liquid, eq. (\ref{dUp=}) implies a modified
Maxwell relation between macroscopic observables
such as $U(t,p)\to U(T,p)= U(T,T_e(T,p),p)$ and $V$.
This solely occurs since $T_e$ is a non-trivial function of 
$T$ and $p$ for the smooth set of experiments under consideration.
Most Maxwell relations involve the entropy, which is difficult to observe. 
The modified Maxwell relation between the observables $U$ and $V$ reads
\BEQ \label{modMaxp}
\frac{\p U}{\p p}\fix_T + p\frac{ \p V}{\p p}\fix_T
+T\frac{\p V}{\p T}\fix_p
=T\frac{\p \I}{\p T}\fix_p\,\frac{\p T_e}{\p p}\fix_T-
T\frac{\p \I}{\p p}\fix_T\,\frac{\p T_e}{\p T}\fix_p+
T_e\frac{\p \I}{ \p p}\fix_T
\EEQ
In equilibrium $T_e=T$, so the right hand side vanishes, and the
standard form is recovered.

In the glass transition region a glass forming liquid exhibits
smeared jumps in the specific heat $C_p=\p (U+pV)/\p T|_p$, 
the expansivity $\alpha=\p \ln V/\ T|_p$
and the compressibility $\kappa=-\p \ln V/\p p|_T$. 
If one forgets about the smearing,
one may consider them as true discontinuities, yielding an analogy
with continuous phase transitions of the classical type.

Following Ehrenfest one may take the derivative of $\Delta
V(T,p_g(T))=0$. The result is 
\BEQ \label{Ehren1p}
\Delta \alpha=\Delta \kappa \frac{\d p_g}{\d T}\EEQ
The conclusion drawn from half a century of research on glass
forming liquids is that this relation is never satisfied
{}~\cite{DaviesJones}\cite{Goldstein}\cite{Jaeckle}~\cite{Angell}.
This has very much hindered progress on a thermodynamical approach.
However, from a theoretical viewpoint it is hard to imagine that
something could go wrong when just taking a derivative.
We have pointed out that this relation is indeed satisfied
automatically~\cite{NEhren}, but it is important say what is
meant by $\kappa$ in the glassy state.

 Previous claims about the violation of the first Ehrenfest relation
can be traced back to the equilibrium thermodynamics idea that there
 is one, ideal $\kappa$,  to be inserted in (\ref{Ehren1p}).
Indeed, investigators always considered cooling curves $V(T,p_i)$
at a set of pressures $p_i$ to determine $\Delta\alpha$ and
$\d p_g/\d T$. However, $\Delta \kappa$ was always determined in
another way, often from measurements of the speed of sound.
In equilibrium such alternative determinations would yield the
same outcome. In glasses this is not the case: the speed of sound is
a short-time process, which only measures short-time relaxation.
Therefore alternative procedures should be avoided, and only
the cooling curves  $V(T,p_i)$ should be used. They constitute
a liquid surface $V_{\rm liquid}(T,p)$ and a glass surface
$V_{\rm glass}(T,p)$ in $(T,p,V)$ space. These surfaces intersect,
and the first Ehrenfest relation is no more than a mathematical
identity about the intersection line of these surfaces.
It is therefore a tautology~\cite{NEhren}, as was also stressed by
McKenna~\cite{McKenna}.

The second Ehrenfest relation follows from differentiating $\Delta
U(T,p_g(T))=0$. The obtained relation will also be satisfied automatically.
However, one then eliminates $\partial U/\partial p$ by means of
the Maxwell relation (\ref{modMaxp}). We obtain
\BEA \label{modEhren2p}
\frac{\Delta C_p}{T_gV}
&=&\Delta\alpha\frac{\d p_g}{\d T}
+\frac{1}{V}\left(1-\frac{\partial T_e}{\partial T}\fix_p\right)
\,\frac{\d \I}{\d T} 
\EEA
The last term, involving  the ``total'' derivative
$\d\I/\d T=\p \I/\p T+(\p\I/\p p)(\d p_g/\d T)$ 
of the configurational  entropy along the glass transition line,
is new.
Its prefactor only vanishes at equilibrium, in which case the
standard Ehrenfest relation is recovered.
The  equality $T_e(T,p_g(T))=T$ implies the useful identity
\BEQ \label{dTedT=1}
\frac{\d T_e}{\d T}=\frac{\p T_e}{\p T}\fix_p+
\frac{\p T_e}{\p p}\fix_T\frac{\d p_g}{\d T}=1 \EEQ

Historically one has defined so-called the Prigogine-Defay ratio
\BEQ
\Pi=\frac{\Delta C_p\Delta\kappa}{TV(\Delta \alpha)^2}
\EEQ
This looks like an equilibrium quantity, and
for equilibrium transitions it should be equal to unity. 
Assuming that at the glass transition a number of
unspecified parameters undergo a phase transition, Davies and Jones
derived that $\Pi\ge 1$~\cite{DaviesJones},
 while DiMarzio showed that in that case the correct value is $\Pi=1$
{}~\cite{DiMarzio}.
In glasses typical experimental values are reported in the range
$2<\Pi<5$. It was therefore generally expected that $\Pi\ge 1$ is
a strict inequality arising from the requirement of mechanical stability.

However, since the first Ehrenfest relation is
satisfied, it holds that~\cite{NEhren}
\BEQ\label{Pip=}
\Pi=\frac{\Delta C_p}{T V\Delta\alpha (\d p_g/\d T)}=1+
\frac{1}{V\Delta \alpha }
\left(1-\frac{\partial T_e}{\partial T}\Bigl|_p \Bigr.\right)
\frac {\d \I}{\d p} \EEQ
Depending on the smooth set of experiments to be performed,
$\d p_g/\d T$ can be small or large: 
$\Pi$ {\it depends on the set of experiments}.
As a result, it can also be below
unity. Rehage-Oels found $\Pi=1.09\approx 1$ at $p=1$
$k\,bar$\cite{RehageOels}, 
using a short-time value for $\kappa$. Reanalyzing their data
we find from (\ref{Pip=}), where the correct $\kappa$ has been inserted,
a value $\Pi=0.77$, which indeed is below unity.
The commonly accepted inequality $\Pi\ge 1$ is based on the
equilibrium assumption of a unique $\kappa$. 
Our theoretical arguments and the
Rehage-Oels data show that such assumptions are incorrect.
  
Further steps involve fluctuation formula and the
fluctuation-dissipation theorem. They are modified outside
equilibrium, and effective temperatures also occur there.
In simple model systems all these effective temperatures
coincide, and this is expected to hold for a subclass of systems
~\cite{Nhammer}\cite{Nlongthermo}.

\section{Black Holes}

Black holes were anticipated by Laplace in 1798. He considered the
gravitational escape problem from a mass $M$. Equating kinetic
$mv^2/2$ to the potential energy $GMm/r$ he observed, for light,
the critical escape radius $R_S=2GM/c^2$. Exactly this radius shows up
in the Schwarzschild solution of the Einstein equations. The
metric for a neutral, non-rotating black hole reads
\BEQ \d s^2=(1-\frac{R_S}{r})\,c^2\,\d t^2-\frac{1}{1-R_S/r}\,\d r^2
-r^2(\, \d \theta^2+\sin^2 \theta \,\d \phi^2)\EEQ
For spherical light waves one has $\d s=\d \theta=\d \phi=0$,
implying a radial speed $\d r/\d t=(1-R_S/r)\,c$, which vanishes
at the `horizon' $r=R_S$.

The connection between black holes and thermodynamics goes back to 
Bekenstein. He introduced in 1973
the notion of black hole entropy, proportional to its area,
in dimensionless units ~\cite{Bekenstein}. 
Since Hawking later fixed the numerical 
prefactor, the result is now called the Bekenstein-Hawking entropy
\BEQ S_{BH}=\frac{k_B A}{4L_P^2}=\frac{\pi R_S^2k_Bc^3}{\hbar G}\EEQ
The presence of $\hbar$ calls for a quantum mechanical 
interpretation, and the quantum evaporation of black holes
was  demonstrated by Hawking ~\cite{Hawking}. 
This  underlined the physical relevance of  Bekenstein's approach. 
The black hole radiates as a black body at Hawking temperature 
\BEQ
\TH=\frac{\hbar G\kappa}{2\pi c^3k_B}=\frac {\hbar c^3}{8\pi GMk_B}\EEQ
where the second equality holds for a non-rotating, neutral black hole.
All possible particles are emitted at this temperature; for large 
black holes, however, $\TH$ is so small, that in practice 
only massless particles  (photons, gravitons,
and possibly neutrino's) are emitted.

A black hole has no hair, i.e. it can be 
 characterized by a few parameters, namely its 
mass $M$, charge $q$ and angular momentum $J$.
This is reminiscent of fluids, that can be characterized by 
temperature and pressure.
From the mass formula for charged, rotating black holes~\cite{Ruffini},
it is known for long that the energy $U=Mc^2$ satisfies ~\cite{BCH}
\BEQ\label{1bhdo}
\d U= \frac{\kappa}{8\pi}\, \d A+\Omega\cdot \d J+\phi\,\d q
\EEQ
where $\kappa$ is the surface tension, $A=4\pi R_S^2$ the area, 
$\Omega$  the horizon's angular velocity, and $\phi$ the
electrostatic potential at the horizon.
This law holds when adding matter to one given black hole,
but also when comparing two different black holes. These two very different
applications suggest a universal validity, and a thermodynamic description. 

These two fundamental results led Bardeen, Carter and Hawking~\cite{BCH}
to formulate ``the four laws of black hole dynamics''.
The zeroth law states that the surface tension $\kappa$ is constant 
at the black hole surface, just like 
the temperature is the same everywhere in an
equilibrium system. The first law is given in eq. (\ref{1bhd}). 
Since the last two terms corresponds to 
work terms, one may write this relation in the suggestive form
\BEQ \label{1bhd}
\d U=\TH\d S_{BH}+\dbarrm W \EEQ
This formulation  is  sometimes referred to as the first law of 
 black holes thermodynamics, but so far it is only a rewriting
of (\ref{1bhdo}).
Bekenstein had already discussed the {\it generalized second law
of black hole dynamics}: the total entropy $S=S_{BH}+S_m$,
where $S_{m}$ is the entropy of the matter outside the black hole,
 cannot decrease: 
\BEQ\label{gen2law}
 \d S\ge 0 \EEQ
The third law concerns ``extremal'' black holes,
the ones that are maximally rotating and/or are maximally charged,
and have $\kappa=0$. The third law of black hole dynamics
says that black holes with $T_H=0$ cannot be reached 
by a finite number of steps~\cite{BCH}\cite{Israel}.

As indicated by their name, the ``four laws of black hole dynamics''
look similar to  the laws of thermodynamics, though a precise
connection was not established. 
Only for a black hole put in a not-too-large box there is equilibrium,
and the standard laws of equilibrium thermodynamics apply
~\cite{Hawking2}\cite{Davies}.
During last 25 years the question has been
to relate these laws to the standard laws of thermodynamics.

From the view point of a condensed matter physicist, the
literature on black hole thermodynamics is somewhat confusing.
First of all, one should define the system for which a
thermodynamic description is to be given. 
A natural choice is to consider as system the black hole
and a ``Gedanken'' sphere around it of, say, 
a hundred times the Schwarzschild radius.
One could also consider the whole 
universe as an isolated container.
Our next objection concerns the non-general formulation of 
the first and second law. 
This has already be discussed in a more general context above. 

Having defined the system, one should determine its entropy. 
For the Gedanken sphere with the black hole in it,
eq. (\ref{1bhd}) applies. The entropy occurring in eq. (\ref{gen2law})
is $ S=S_{BH}+S_m^{\rm Gs}$.
The latter is the entropy of the cosmic background matter 
outside the black hole but inside the Gedanken sphere, and 
expected to scale with the sphere's volume. 
The  radiation generated by the hole will quickly leave the system
and go to the heat bath around it; this is described by a $\dbarrm Q<0$. 

If, on the other hand, the whole universe is considered as system,
then $\dbarrm Q=0$. If no work is done, this implies that $\d U=0$, 
saying that  energy radiated from the hole is still inside 
the system. In that case eq. (\ref{1bhd}) does not describe 
the change of the system's energy, it only says something about the
black hole. The total entropy is now $S=S_{BH}+S_m$, and 
the second law of thermodynamics indeed says that $\d S\ge 0$.

We conclude that eq. (\ref{1bhd}) and (\ref{gen2law}) are both valid,
though they should not be
applied simultaneously. They refer to different situations, that is 
to say, to different time scales. When only the black hole
and its Gedanken sphere are considered, this describes
the radiation emitted in per unit time.
When considering the change in entropy of the whole universe,
one tacitly assumes time scales so 
large that all information on the emitted radiation has been lost.
Notice that the negative curvature of the universe, leading to
exponential divergence of nearby trajectories, can 
establish this loss of information even though 
the emitted photons are hardly scattered.

A final, severe, objection against the current formulation of
thermodynamics for black holes is: {\it  what is the heath bath?}
When working with one temperature, $T_H$, this is by definition 
the bath temperature, and normally also the temperature of the object.
This can only apply to a black hole in
equilibrium with its own Hawking radiation, which is an unstable
and thus unphysical situation; such an approach 
can also not deal with black holes of different size.
Physically there is one, and only one choice for the bath: for a
black hole far away from matter, the heat bath is the cosmic microwave 
background radiation, that presently has temperature $\Tbg=2.73\,K$
 (we neglect here the small CMB-fluctuations).
So the actual problem deals with a system of which the dynamics
``lives'' at a second temperature, namely $\TH$.
As for glasses, this is a two-temperature situation,
in agreement with the time scale argument: Black holes heavier than 
$10^{-18}M_\odot=10^{15}\,g$ 
need more time to evaporate than the present age of the universe.
For them the evaporation process, as seen by far-away observers,
is so slow,  that equilibration of the cosmic background radiation 
is a fast process.

The slow evaporation processes occur at the Hawking temperature 
and have entropy $S_{BH}$. 
The entropy of the background radiation outside 
the back hole but inside the Gedanken sphere is very small.
Moreover, the emitted radiation will immediately 
leave this sphere and there is no `back reaction', so 
$\d S_m^{\rm Gs}= 0$. This simplifies 
eq. (\ref{dQ=}) to
\BEQ \label{2lawbh1}
\dbarrm Q=T_H \d S_{BH}
\EEQ
The standard first law $\d U=\dbarrm Q+\dbarrm W$ 
thus indeed reproduces (\ref{1bhd}), 
notwithstanding its non-equilibrium thermodynamic
nature with $T_{BH}\neq \Tbg$.
According to the second law (\ref{gen2law}), $S_{BH}$ has to satisfy 
\BEQ\label{2lawbh} 
(\Tbg-T_H)\,\d S_{BH} \ge 0
\EEQ 
Hawking radiation leads to $\d S_{BH}<0$. Eq. (\ref{2lawbh}) is thus 
satisfied as long as $T_H>\Tbg$, but not below that. One might
think that $\Tbg$ plays no physical role whatsoever, and only shows up
as determinator in the second law.
However, the real point is that we not yet
considered absorption of background radiation by the black hole.
The  absorption rate will be proportional to the
area times the energy density, i.e., $\sim M^2\Tbg^4$.  
The quantum absorption process 
 is simply the time-reversed evaporation process.  
For non-rotating,
neutral holes Hawking radiation  leads to a mass loss 
\BEQ
\dot M=-\alpha_{em}\frac{\hbar c^4}{G^2M^2}
\to \dot T_H=\frac{(8\pi)^3 \alpha_{em}G k_B^3}{\hbar^2 c^5}\TH^4
\EEQ
The dimensionless constant $\alpha_{em}$ depends on the type of particles
present, and their absorption probabilities, called
``oscillator strengths'' in solid state systems.
$\TH$ enters through the Bose-Einstein occupation numbers (for bosons,
in particular photons) or Fermi-Dirac occupation numbers (for fermions). 
For an uncharged, non-rotating black hole Page finds 
$\alpha=5.246\times 10^{-4}$ in the high-frequency limit, and
$0.181\times 10^{-4}$ in the low frequency limit~\cite{Page}.
For absorption by the black hole of a photon (or a particle) from the 
cosmic background, the time-reversed problem shows up. 
It thus holds that $\alpha_{abs}(T)=\alpha_{em}(T)$,
no matter the character of the particle content; 
for simplicity we shall now replace both by a constant. 
The only difference between the
two situations is the temperature occurring in the occupation numbers: 
for Hawking emission it is $T_H$,
while for cosmic background absorption it is $\Tbg$. 
The combined processes of Hawking emission and background photon
absorption thus yields for a neutral, non-rotation black hole ~\cite{Zurek}
\BEQ\label{balance}
\dot T_H=\frac{ (8\pi)^3 \alpha G k_B^3 }{\hbar^2c^5}\,(\TH^4-\Tbg^4)
\EEQ
It exhibits an instability at $\TH=\Tbg$,
related to the fact that the ``Bekenstein-Hawking''
specific heat $C_{\rm BH}=\d U/\d \TH$ is negative. 
If there is equilibrium, and $\Tbg$ is changed a little, 
then $T_H$ branches away from it.

There are two regimes. In the ``classical''
regime $\TH<\Tbg$ the black hole absorbs more radiation than is emits.
Its entropy will increase, and $\dbarrm Q=T_H\d S_{BH}>0$, 
but this is still in accord with the second law (\ref{2lawbh}).
In the ``quantum'' regime $T_H>\Tbg$ the black hole emits 
more than it absorbs. Now it holds that $\d S_{BH}<0$, confirming
again that heat flows from high to low temperature.

In analogy with glasses, one can define the {\it apparent specific heat}
$C=\p U/\p \Tbg =\dot U/\dot \Tbg$.
For black holes this object is less natural because 
the background temperature cannot be changed by hand. 
However, $C$ does have a meaning in
our expanding universe. Due to the decrease of $\Tbg$, there will be
less and less background energy to be absorbed. 
Eq. (\ref{balance}) tells us that,
provided  $\Tbg$ decays quicker than $t^{-1/3}$, 
a black hole will reach its maximal size  at some moment $t=t_0$ where
the temperatures match, $\TH=\Tbg=T_0$; from then on it will shrink,
and its Hawking temperature rises. In our matter dominated Universe
one has $ \Tbg\sim t^{-2/3}$, so finally the black hole will evaporate. 
Around $t_0$ the apparent specific heat 
takes a form independent of $\dot\Tbg$, viz.
$C=k_B(t-t_0)/\tau$, with characteristic time scale the quantum time
$\tau=\hbar/[(16\pi)^2\alpha k_B T_0]$.
In the classical regime ($t<t_0$) $C$ is negative, while
in the quantum regime it is positive.

The third law of black hole dynamics is related to extremal black 
holes, that have $T_H=0$. The third law of thermodynamics, however, 
concerns the vanishing of the 
 entropy for $\Tbg\to 0$. We have seen already that 
finally all black holes evaporate,  thereby 
lowering their configurational entropy very much, in accord
with Planck's third law.  Notice that this has nothing to do with the
third law of black hole dynamics.
What happens ultimately with the black hole
has been the focus of studies by 't Hooft ~\cite{tHooft}.

The entropy change of the universe, now assumed to be a thermal
bath, is found as for  black body radiation~\cite{Zurek}.
Changes of the energy and entropy of the black hole surroundings
due to absorption and emission by the black hole are
$\dot U_m\sim\alpha(\TH^4-\Tbg^4)$, $\dot S_m\sim(4/3)(\TH^3-\Tbg^3)$,
which leads to
\BEQ\label{dSdIGs}
\frac{\d S_m}{\d S_{BH}}=\frac{\TH\d S_m}{\d U}=-\frac{\TH\d S_m}{\d U_m}
=-\frac{4\TH(\TH^3-\Tbg^3)}
{3(\TH^4-\Tbg^4)}\EEQ
This yields the entropy production $\dot S=\dot S_m+\dot S_{BH}$ 
\BEQ\label{SprodGs}
\dot S=\frac {(8\pi)^2\alpha k_B^2}{3\pi\hbar}\,
\frac{(\TH^2+2\TH\Tbg+3\Tbg^2)(\TH-\Tbg)^2}{\TH^3}
\EEQ

Next we consider the whole universe as our system, so
the entropy of the universe $S_m$ has to be taken into account.
While $S=S_m+S_{BH}$ is the total entropy, eq. (\ref{dQ=})
becomes $\dbarrm Q=\Tbg\d S_m+\TH\d S_{BH}$. As $\dbarrm Q=0$, 
the second law again leads to (\ref{2lawbh}), 
but with different entropy production. For the present case we find the
new result 
\BEQ\label{Sdotuniv}
\dot S=\frac {(8\pi)^2\alpha k_B^2}{\hbar}\,
\frac{(\TH^4-\Tbg^4)(\TH-\Tbg)}{\TH^3\Tbg}
\EEQ
This expression exceeds eq. (\ref{SprodGs}), which referred to
radiation that was still located near the black hole.
The difference is due to loss of information on the emitted radiation.

Our approach thus incorporates the known properties of dynamics,
and shows how the generalized second law comes into the play. 
Both the formation and evaporation of
black holes leads to an increase of the entropy of the whole universe.
Our picture involves the standard zero-entropy formulation of the third law
of thermodynamics, thus putting aside the third law of black hole mechanics.
To the best of our knowledge, there is no contradiction 
with the occurrence of negative specific heats.

\end{document}